\documentstyle[12pt]{ioplppt}
\input epsf.tex

\begin{document}

\jl{1}

\title{Aperiodic spin chain in the mean-field 
approximation}[Aperiodic spin chain in the mean-field approximation]
\author{Pierre Emmanuel Berche and Bertrand Berche\ftnote{1}{To whom 
correspondence should be
addressed}}
 
\address{Laboratoire de Physique des Mat\'eriaux\ftnote{3}{Unit\'e de
Recherche Associ\'ee au CNRS No 155},  Universit\'e Henri Poincar\'e, Nancy~1,
BP 239, F--54506 Vand\oe uvre les Nancy Cedex, France}

\date{\today}

\begin{abstract}
Surface and bulk critical properties of an aperiodic spin chain  are 
investigated
 in the framework of the $\phi^4$ phenomenological Ginzburg-Landau
theory. According to Luck's criterion,  the mean field correlation length
exponent $\nu=1/2$ leads to a marginal behaviour when the wandering exponent of 
the sequence is
$\omega=-1$. This is the case of the Fibonacci sequence that we consider here.  
We calculate the bulk and surface critical
exponents  for the magnetizations, critical isotherms, susceptibilities and 
specific heats. These
exponents continuously vary with the amplitude of the perturbation. 
Hyperscaling 
relations are used in order to obtain an estimate of the upper critical 
dimension for this system.
\end{abstract} 

\maketitle

\section{Introduction}
\label{sec:intro}

The discovery of quasicrystals~\cite{schechtman84}  has focused considerable
interest on quasiperiodic or, more
generally, aperiodic systems~\cite{henley87a}. 
In the field  of critical 
phenomena, due to their intermediate situation between periodic and random
systems,  aperiodic models have been intensively studied (for a review, 
see~\cite{grimm96}).
Furthermore, aperiodic multilayers are experimentally feasible  and should
build a  new class of artificial structures exhibiting interesting bulk and 
surface properties.
Although aperiodic superlattices have already been worked out by molecular beam
epitaxy~\cite{majkrzak91}, nothing has been done experimentally up to now from
the point of view of critical phenomena.
In the perspective of possible future experimental
studies in this context, it seems an interesting and challenging problem to 
complete our
understanding through a mean field theory approach. 
 Surface critical behaviour has indeed been intensively investigated on
the basis of the Ginzburg-Landau theory~\cite{ginzburg50} in the
seventies~\cite{mills71}.
This led to a classification of the transitions which may occur at the
surface and to the derivation of scaling laws between surface and bulk critical
exponents~\cite{bray77a} (for a review,
see~\cite{binder83}). These early papers are known as an important stage in
the further developments of surface critical phenomena.

Seen from the side of critical phenomena,  the universal behaviour of 
aperiodically perturbed
systems is now well understood since Luck proposed a
relevance-irrelevance criterion~\cite{luck93a,luck93b}.
The characteristic length scale in a critical system is given by the 
correlation 
length and as
in 
 the Harris criterion for random systems~\cite{harris74}, the
strength of the fluctuations of the couplings on  this scale determines the 
critical behaviour.
An aperiodic perturbation can thus be relevant, marginal or irrelevant, 
depending on the
sign of a crossover exponent involving the correlation length exponent $\nu$ of 
the
unperturbed system  and the wandering exponent $\omega$ which governs the 
size-dependence
of the fluctuations of the aperiodic couplings~\cite{queffelec87}. In the light
of this criterion, the results obtained in early papers, mainly concentrated on
the
Fibonacci~\cite{igloi88}
and the Thue-Morse~\cite{doria89} sequences, found a consistent
explanation,
 since, resulting from 
the bounded fluctuations, a critical
behaviour which  resembles the periodic case was found for the Ising model in 
two dimensions.

In the last years, much progress have been made in the understanding of the
properties of marginal and relevant aperiodically perturbed systems. Exact 
results for the 
$2d$ layered Ising model and the quantum Ising chain   have  been
obtained with irrelevant, marginal and relevant aperiodic
perturbations~\cite{lin92a,iglotu94}.
The critical behaviour is in agreement with Luck's criterion leading to 
essential singularities
or first-order surface transition when the perturbation is relevant and power 
laws with
continuously  varying exponents in the marginal situation with logarithmically 
diverging
fluctuations.  A strongly anisotropic behaviour has been recognized in this 
latter 
situation~\cite{berche95,berche96}. Marginal surface perturbations have also 
been
studied with the Fredholm sequence~\cite{karevski95a} and conformal aspects 
have 
been
discussed~\cite{grimm94}.

In the present paper, we continue our study of marginal
sequences. The case of the Fibonacci sequence, which leads to irrelevant 
behaviour in the
Ising model, should exhibit non universal properties within  mean field 
approach according to
the Luck criterion and it has not yet been studied in this context. The article 
is organized as
follows: in \sref{sec:GL},   we present  the phenomenological  Ginzburg-Landau
theory on a discrete lattice with a perturbation following a Fibonacci sequence
and we summarize the scaling arguments leading to  Luck's criterion, then we
discuss the definitions of both bulk and surface thermodynamic quantities. We 
consider magnetic properties in \sref{sec:mag}. Both bulk and surface 
quantities
are computed numerically, leading to the values of the corresponding critical
exponents. In \sref{sec:therm}, we discuss the thermal properties and
eventually  in \sref{sec:concl}, we discuss  the upper critical dimension of 
the model.

\section{Discrete Ginzburg-Landau equations for a Fibonacci aperiodic
perturbation}\label{sec:GL}

\subsection{Landau expansion and equation of state on a one-dimensional 
lattice}

Let us first review briefly the essentials of the Ginzburg-Landau theory
formulated on a discrete lattice. We consider a one-dimensional lattice of  $L$
sites  with a lattice spacing $\ell$ and free boundary conditions. The critical 
behaviour would be
the same  as in  a $d-$dimensional plate of thickness $L\ell$ with
translational invariance along the $d-1$ directions perpendicular to the chain
and extreme axial anisotropy which forces the magnetic moments to keep a
constant direction in the plane of the plate. We investigate the critical
properties of an aperiodically distributed perturbation within the framework of
a $\phi^4$ phenomenological Landau theory~\cite{landau}. The underlying
assumption in this approach is based on the
 following expansion of the bulk
free energy density 
 \begin{equation}
f_b\{\phi_j\}={1\over 2}\mu_j\phi_j^2+{1\over 4}g\phi_j^4-H\phi_j+{1\over
2}c\left({\phi_{j+1}-\phi_j\over\ell}\right)^2,\label{eq-1} \end{equation}
where the aperiodic perturbation of the coupling constants is determined by a 
two-digits
substitution rule and enters the $\phi^2$ term only. A dimensional analysis 
indeed shows that
the deviation from the critical temperature, $\mu$, is  the relevant scaling 
field which has to
be modified by the perturbation. The free energy of the whole chain is thus 
given by
\begin{equation}
F[\phi_j]=\sum_jf_b\{\phi_j\},
\label{eq-4}\end{equation}
and the spatial distribution of order parameter  satisfies the usual functional 
minimization:
\begin{equation}
\delta F[\phi_j]=F[\phi_j+\delta\phi_j]-F[\phi_j]=0.
\label{eq-5}\end{equation} 
One then obtains the coupled discrete Ginzburg-Landau equations:
\begin{equation}
\mu_j\phi_j+g\phi_j^3-H-{c\over\ell^2}(\phi_{j+1}-2\phi_j+\phi_{j-1})=0.
\label{eq-6}\end{equation}
The coefficients $\mu_j$ depend on the site location and are written as
\begin{equation}
\mu_j=k_BT-(zJ-Rf_j)=a_0\left(1-{1\over\theta}+rf_j\right),
\label{eq-7}\end{equation}
where $J$ is the exchange coupling between neighbour sites in the homogeneous 
system, $z$ the
lattice coordination and $f_j$ the aperiodically distributed sequence of $0$ 
and 
$1$.
The prefactor $a_0=k_BT$ is essentially constant in  the vicinity of the
critical point, and the temperature $\theta$ is normalized relatively to the
 unperturbed system critical temperature: 
$\theta=k_BT/zJ$. In the following, we will also use the notation 
$\mu=1-1/\theta$. In
order to obtain a dimensionless equation, let us define 
$\phi_j=m_j\sqrt{a_0/g}$
 leading to the following non-linear equations for the $m_j$'s:
\begin{equation}
(\mu+rf_j)m_j+m_j^3-h-(m_{j+1}-2m_j+m_{j-1})=0,
\label{eq-8}\end{equation}
with boundary conditions
\numparts
\begin{eqnarray}
&(\mu+rf_1)m_1+m_1^3-h-(m_{2}-2m_1)=0,\\
&(\mu+rf_L)m_L+m_L^3-h-(-2m_L+m_{L-1})=0.
\end{eqnarray}
\label{eq-9}\endnumparts 
Here,
the lengths are measured in units $\ell=\sqrt{c/a_0}$ and $h=H\sqrt{g/a_0^3}$ 
is
a reduced magnetic field.

One can point out the absence of specific surface term 
in
the free energy density.  The surface equations for the order parameter profile  
 simply keep the
bulk form with the boundary conditions $m_0=m_{L+1}=0$ and our study will only 
concern ordinary
surface transitions~\cite{binder83}.

 \subsection{Fibonacci perturbation and Luck's criterion}

The Fibonacci perturbation considered below may be defined as a two digits 
substitution
sequence which follows from the inflation rule 
\begin{equation}
0\rightarrow S(0)=01,\quad
1\rightarrow S(1)=0,
\label{eq-11}\end{equation} 
leading, by iterated application of the rule on the initial word $0$, to 
successive words of
increasing lengths: 
\begin{equation}
\begin{array}{llllllll}
0&&&&&&&\\
0&1&&&&&&\\
0&1&0&&&&&\\
0&1&0&0&1&&&\\
0&1&0&0&1&0&1&0\\
.&.&.&&&&&\\
\end{array}
\label{eq-12}\end{equation}
It is now well known that most of the properties of such a sequence can be
characterized by a substitution matrix whose elements $M_{ij}$ are given by the
number $n_i^{S(j)}$ of digits of type $i$ in the substitution
$S(j)$~\cite{luck93a,queffelec87}. In the case of the Fibonacci sequence, this
yields   \begin{equation}
\mat{M}=
\left(
\begin{array}{cc}
n_0^{S(0)}&n_0^{S(1)}\\
n_1^{S(0)}&n_1^{S(1)}\\
\end{array}
\right) =
\left(
\begin{array}{cc}
1&1\\
1&0\\\end{array}
\right).
\label{eq-13}\end{equation} 
The largest eigenvalue of the
substitution matrix
 is given by the
golden mean $\Lambda_1={1+\sqrt{5}\over 2}$ and is related to the length of the
sequence after $n$ iterations, $L_n\sim\Lambda_1^n$, while the second 
eigenvalue
$\Lambda_2=-1/\Lambda_1$ governs the behaviour of the cumulated deviation
from the asymptotic density of modified couplings
$\rho_\infty=1-{2\over\sqrt 5+1}$: 
\begin{equation}
\sum_{j=1}^L(f_j-\bar f)=n_L-\rho_\infty
L\sim\mid\Lambda_2\mid^n\sim (\Lambda_1^\omega)^n,
\label{eq-14}\end{equation} 
where we have introduced
the sum $n_L=\sum_{j=1}^Lf_j$  and the wandering exponent \begin{equation}
\omega={\ln\mid\Lambda_2\mid\over\ln\Lambda_1}=-1.
\label{eq-15}\end{equation}

When the scaling field $\mu$ is perturbed as considered in the previous 
section, 
\begin{equation}
\mu_j=a_0\left(\mu+rf_j\right),
\label{eq-16}\end{equation}
the cumulated deviation of the couplings from the average at a length scale $L$
\begin{equation}
\overline{\delta\mu}(L)={1\over L}\sum_{j=1}^L(\mu_j-\bar\mu)=
{1\over L}a_0r(n_L-\rho_\infty L)
\label{eq-17}\end{equation}
behaves with  a size power law:
\begin{equation}
\overline{\delta\mu}(L)\sim L^{\omega-1},
\label{eq-18}\end{equation}
and induces a shift in the critical temperature $\overline{\delta
t}\sim\xi^{\omega -1}$ to be compared with the deviation $t$ from the critical 
temperature:
\begin{equation}
{\overline{\delta t}\over t}\sim t^{-(\nu (\omega -1)+1)}.
\label{eq-41}
\end{equation}
This defines the crossover exponent $\phi=\nu (\omega -1)+1$. When $\phi=0$, 
the
perturbation is marginal: it remains unchanged under a renormalization 
transformation,  and the system is thus
governed by a new perturbation-dependent fixed point.

A perturbation of the
parameters $g$ or $c$ entering the Landau expansion \eref{eq-1} would be 
irrelevant.

\subsection{Bulk and surface thermodynamic quantities}

In the following, we discuss both bulk and surface critical exponents and 
scaling functions. We
deal with the surface and boundary magnetizations $m_s$ and $m_1$, surface and 
boundary
susceptibilities $\chi_s$ and $\chi_1$, and surface specific heat $C_s$. All
these quantities can be expressed as  derivatives of  the surface free energy
density $f_s$~\cite{binder83} (see \tref{table4}).

\begin{table}
\caption{Bulk and surface thermodynamic quantities in terms of the 
bulk $f_b$ and surface
$f_s$ free energy densities. $h$ and $h_1$ are bulk and surface 
magnetic fields respectively
and $t$ is the reduced temperature. }\footnotesize\rm
\begin{tabular}{@{}llllllll} \br
\centre{2}{magnetization}&&\centre{2}{susceptibility}&&\centre{2}{specific 
heat}\\
\crule{2}&\quad&\crule{2}&\quad&\crule{2}\\
bulk&surface&\quad&bulk&surface&\quad&bulk&surface\\ \mr
$m_b=-{\partial f_b\over\partial h}$ & $m_s=-{\partial f_s\over\partial h}$ 
&\quad&
$\chi_b=-{\partial^2 f_b\over\partial h^2}$ & $\chi_s=-{\partial^2
f_s\over\partial h^2}$ &\quad& $C_b=-{\partial^2 f_b\over\partial t^2}$ & 
$C_s=-{\partial^2
f_s\over\partial t^2}$ \\ & $m_1=-{\partial f_s\over\partial h_1}$ &\quad& &
$\chi_1=-{\partial^2 f_s\over\partial h\partial h_1}$ &\quad& & \\ & &
&\quad& $\chi_{11}=-{\partial^2 f_s\over\partial h_1^2}$ &\quad& & \\
\br\end{tabular} \label{table4} \end{table}

\begin{figure}
\epsfxsize=11cm
\begin{center}
\mbox{\epsfbox{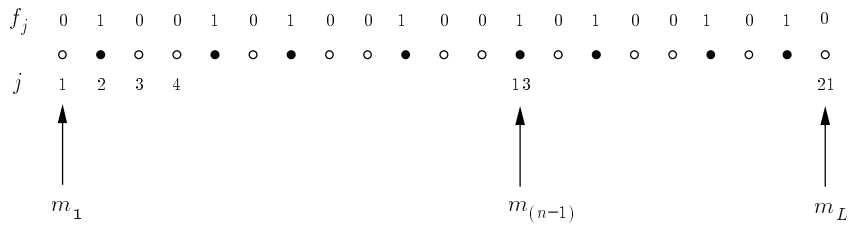}}
\end{center}
\vskip 0mm
\caption{Fibonacci chain of 21 sites.  The local bulk
magnetization, for a chain of $L_n$ sites obtained after $n$ iterations of
the substitution rule is computed on the site $L_{n-1}$, here site 13.} 
\label{fig2}  
\end{figure}

While there is no special attention to pay to these definitions in a 
homogeneous 
system, they
have to be carefully rewritten in the perturbed model that we consider here. 
First of all, we
shall focus on local quantities such as the boundary magnetization $m_1$ or the 
local bulk
magnetization $m_{(n-1)}$, defined,   for a chain of size $L_n$ obtained after 
$n$
substitutions, by the order parameter at position $L_{n-1}$. This definition 
leads to equivalent sites
for different chain sizes (see \fref{fig2}).

\begin{figure}
\epsfxsize=11cm
\begin{center}
\mbox{\epsfbox{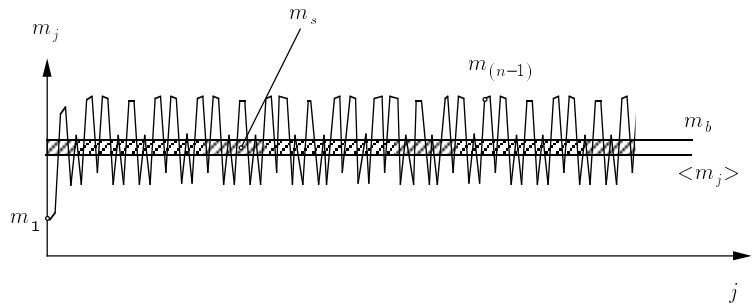}}
\end{center}
\vskip 0mm
\caption{Typical shape of the order parameter profile for a perturbed system, 
showing the
boundary  and local bulk  magnetizations $m_1$ and $m_{(n-1)}$, and the average 
values $ m_b$ and
$\langle m_j\rangle$} 
\label{fig10}   \end{figure}

In addition to these local quantities, one may also calculate both surface and 
mean bulk
magnetizations ($m_s$ and $m_b$ respectively),
 which should  be interesting from an  experimental
point of view since any experimental device would average any measurement over 
a
large region compared to microscopic scale.
 In order to keep symmetric sites with respect to the middle of the chain, and 
to avoid
surface effects, the mean  bulk magnetization $m_b$ is defined by averaging
over  $L_{n-2}$ sites around the middle for a chain of size $L_n$.
\begin{equation}
m_b={1\over L_{n-2}}\sum_{j\in L_{n-2}}m_j.\label{eq-100}
\end{equation}
We checked numerically that one recovers the same average as for a chain of 
size 
$L_{n-2}$ with
periodic boundary conditions. 
Following Binder~\cite{binder83}, for a film of size $L_n$ with two free 
surfaces, the surface
magnetization is then defined by the deviation of the average magnetization 
$\langle
m_j\rangle$ over the whole chain from the bulk mean value: 
\begin{equation}
m_s={1\over 2}\left( m_b-{1\over L_n}\sum_{j=1}^{L_n}m_j\right).
\label{eq-36}\end{equation}
A graphical description can be found in \fref{fig10}. 

In the following, we shall use  brackets for the
averages over the finite system, taking thus surface effects into account.
In the same way,  the bulk free energy density in \tref{table4} has to be 
understood as:
\begin{equation}
 f_b={1\over L_{n-2}}\sum_{j\in L_{n-2}} f_b\{m_j\}
\label{eq-40}\end{equation}
while the surface free energy density $f_s$
 is defined as  the excess  from
the average bulk free energy
\begin{equation}
F=\sum_{j=1}^{L_n}f_b\{m_j\}=L_n\langle f_b\rangle=L_n f_b+2f_s.
\label{eq-37}\end{equation}

 \section{Magnetic properties}
\label{sec:mag}

\subsection{Order parameter profile and critical temperature} 

\begin{figure}
\epsfxsize=11cm
\begin{center}
\mbox{\epsfbox{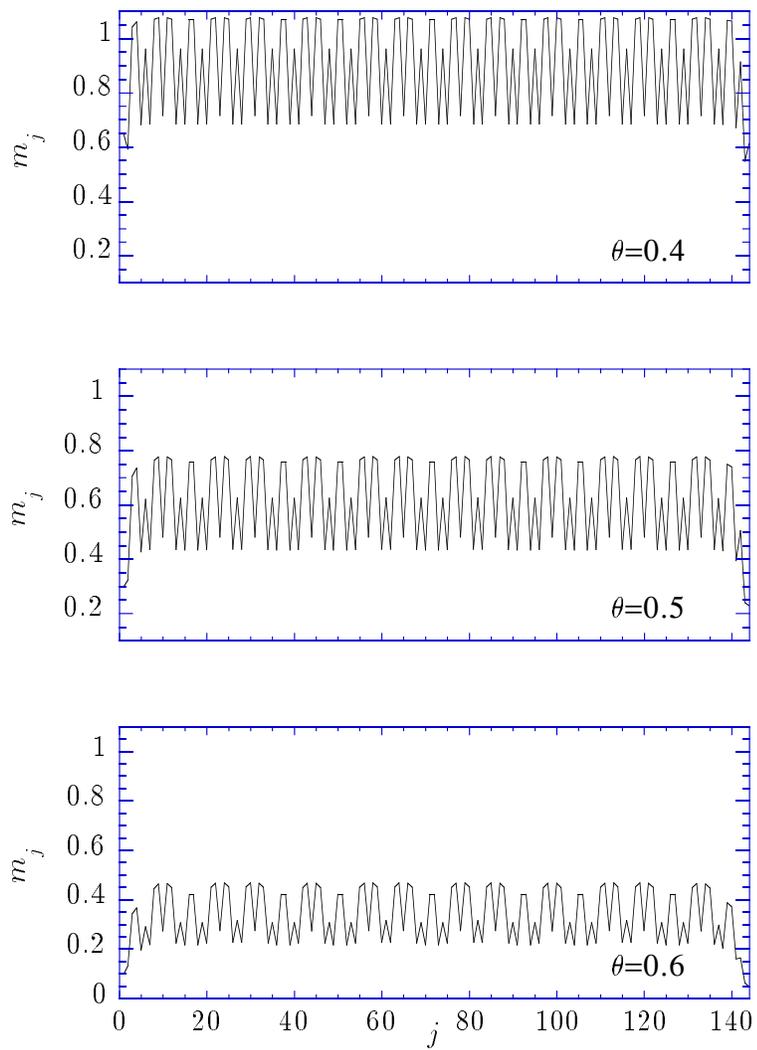}}
\end{center}
\vskip -0mm
\caption{Order parameter profiles for a perturbation $r=2$ and three values of
the temperature below the critical point. The size of the chain is $L=144$.} 
\label{fig1}  
\end{figure}

The order parameter profile is determined numerically by a Newton-Raphson
method, starting with arbitrary values for the initial trial profile ${m_j}$.
\Eref{eq-8} provides a system of $L$ coupled non-linear equations
\begin{equation}
G_i(m_1,m_2,\dots,m_L)=0,\quad i=1,2,\dots ,L
\label{eq-29}
\end{equation}
for the components of the vector $\vec m=(m_1,m_2,\dots,m_L)$, which can be 
expanded  in a first
order Taylor series: \begin{equation}
G_i(\vec m+\delta \vec m)=G_i(\vec m)+\sum_{j=1}^L{\partial G_i\over\partial 
m_j}\delta
m_j+O(\delta\vec m^2).
\label{eq-30}
\end{equation}
A set of linear equations follows for the corrections $\delta\vec m$
\begin{equation}
\sum_{j=1}^L{\partial G_i\over\partial m_j}\delta
m_j=-G_i(\vec m)
\label{eq-31}
\end{equation}
which moves each function $G_i$ closer to zero simultaneously.
This technique is known to provide a  fast convergence towards the exact
solution. Typical examples of the profile obtained for the Fibonacci
perturbation are shown on~\fref{fig1}.

The magnetization profile decreases as the temperature is increased and 
vanishes
for some size-dependent effective value of the critical temperature
$\mu_c(L)=1-(\theta_c(L))^{-1}$. This value may be obtained through a recursion
relation deduced from the equation of state. In the high temperature phase, 
when 
$h=0$,
equation \eref{eq-8}  can  be rewritten as a homogeneous system of linear 
equations:
\begin{equation}
\mat{G}\vec m
=
\left(
\begin{array}{cccccc}
\alpha_1& -1&0&0&\dots&0\\
-1&\alpha_2&-1&0&\dots&0\\
0&-1&\alpha_3&-1&\ddots&\vdots\\
\vdots&\vdots&\ddots&\alpha_j&\ddots&\vdots\\
0&\dots&0&0&-1&\alpha_L\\
\end{array}
\right)
\left(
\begin{array}{c}
m_1\\
m_2\\
\vdots\\
\vdots\\
m_L\\
\end{array}
\right)
=0,
\label{eq-20}\end{equation}
where $\alpha_j=2+\mu+rf_j$.
If the determinant $D_L(\mu)=\hbox{\rm Det}\ \mat{G}(\mu)$ is not
vanishing, the null vector ${\vec m}=\vec 0$ provides the satisfying unique 
solution
for the high temperature phase. The critical temperature is then  defined
by the limiting value $\mu_c(L)$ which allows a non-vanishing solution for 
${\vec m}$,
i.e. $D_L(\mu_c)=0$. Because of the tridiagonal structure of the determinant,
the following recursion relation holds, for any value of $\mu$:
\numparts
\begin{eqnarray}
D_L(\mu)&=&\alpha_LD_{L-1}(\mu)-D_{L-2}(\mu),\\
D_0(\mu)&=&1,\\
D_1(\mu)&=&\alpha_1.
\end{eqnarray}\label{eq-23}
\endnumparts

Thus we can obtain $\mu_c(L)$ for different sizes $L$ from 144 to 46368 and 
estimate the
asymptotic critical point by an extrapolation to infinite size. This technique 
allows a
determination of the critical temperature with an absolute accuracy in the 
range 
$10^{-7}$ to
$10^{-9}$ depending on the value of the amplitude $r$.

\subsection{Surface and bulk spontaneous magnetization behaviours}

\begin{figure}
\epsfxsize=11cm
\begin{center}
\mbox{\epsfbox{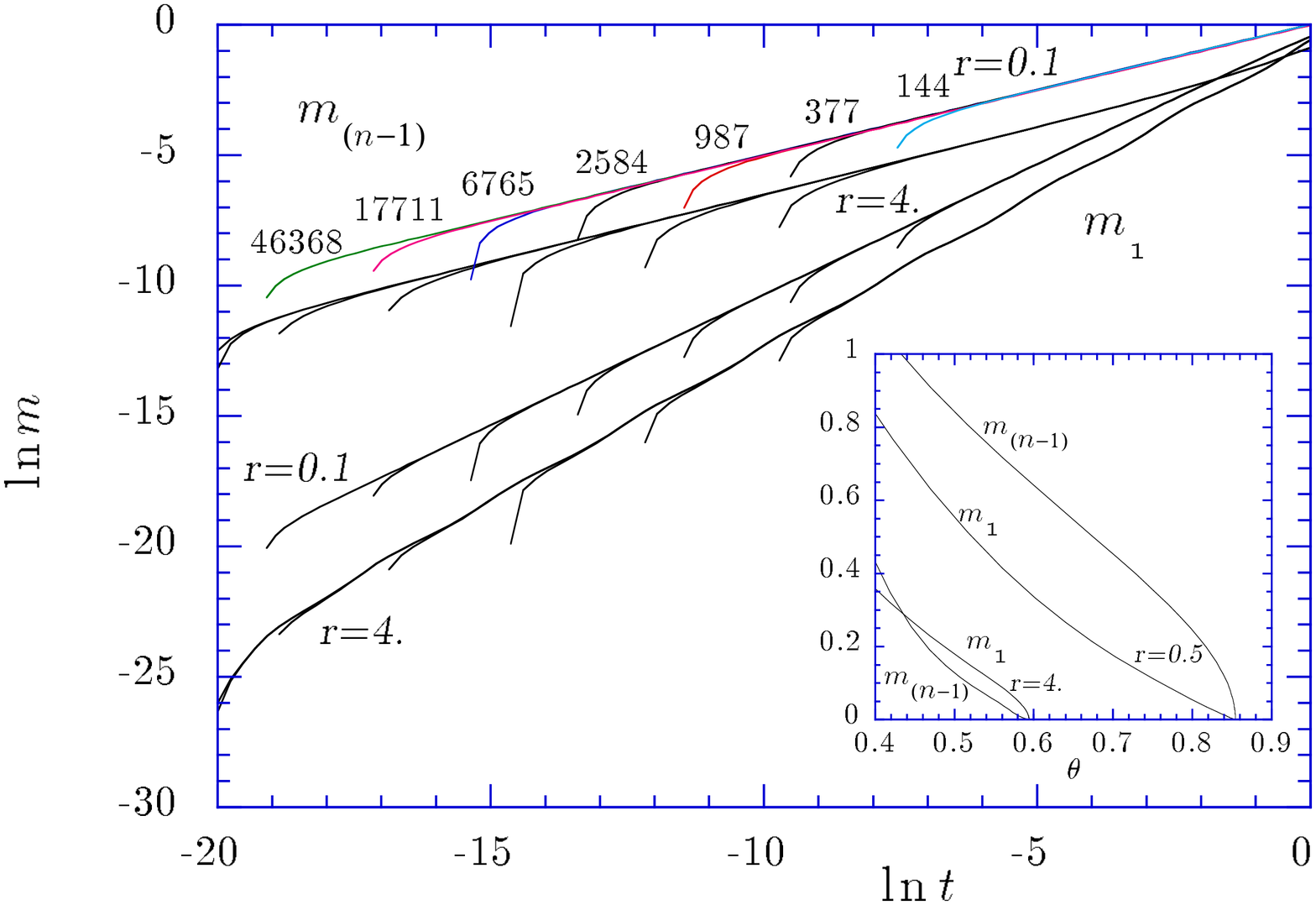}}
\end{center}
\vskip -0mm
\caption{Log-Log plot of the bulk and boundary magnetization  v.s. the reduced
temperature $t=\mu_c-\mu$ for two values of the aperiodic amplitude $r$ and for
different sizes of the chain from 144 to 46368. Finite-size effects occur when
the curves deviate from the asymptotic straight line.  The insert shows the 
behaviour
of the magnetization with the temperature.} 
\label{fig3}  
\end{figure}

The boundary magnetization $m_1$ vanishes at the same temperature than
 the profile itself. First of all, the influence
of finite size effects~\cite{barber83} has to be studied. This is done by the
determination of the profiles for different chains of lengths given
by the successive sizes of the Fibonacci sequence $L=1,2,3,5,8,13,21,34\dots$ 
The boundary and bulk magnetization in zero magnetic field are shown 
on~\fref{fig3}
on a log-log scale. 

The finite size effects appear in the deviation from the straight line 
asymptotic behaviour.
These effects are not too sensitive, as it can  be underscored by considering 
the
deviation of the curve for a size $L=17711$, which occurs around
$t=\mu_c-\mu\simeq 10^{-7}$, i.e.  very close to the critical point. 

The expected marginal
behaviour is furthermore indicated by the variation of the  slopes with the 
aperiodic modulation
amplitude $r$ and is more noticeable for the boundary magnetization than in the 
case of the bulk.

\begin{figure}
\epsfxsize=11cm
\begin{center}
\mbox{\epsfbox{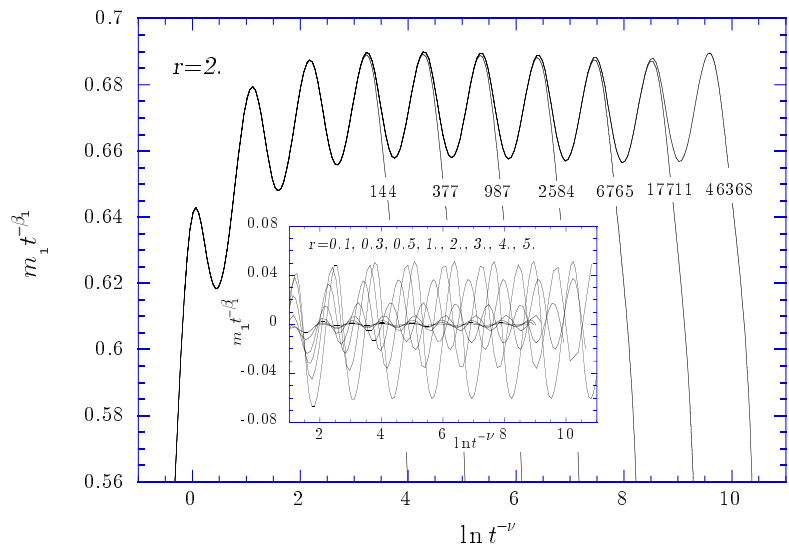}}
\end{center}
\vskip -0mm
\caption{Periodic oscillations of the rescaled boundary magnetization
$m_1t^{-\beta_1}$  v.s. $\ln t^{-\nu}$. The deviation
from the oscillating behaviour for large values of the correlation length 
$t^{-\nu}$ is due to
finite-size effects. The insert shows the oscillations of the rescaled boundary
magnetization  for
different values of $r$ after substraction of a constant amplitude.} 
\label{fig4}  
\end{figure}

A more detailed inspection of
these curves also shows oscillations resulting from the discrete scale
invariance~\cite{jona75} of the system and the asymptotic
magnetization can thus be written 
\begin{equation}
m(t)=t^\beta\tilde m(t^{-\nu})
\label{eq-25}\end{equation}
where $\tilde m(t^{-\nu})$ is  a log-periodic scaling function of its argument.
We make use of this oscillating behaviour to obtain a more precise 
determination 
of
the critical temperature (in the range $10^{-11}$ to $10^{-12}$) and of the 
values 
of the bulk and
surface exponents by plotting the rescaled magnetization $mt^{-\beta}$ as a 
function of $\ln
t^{-\nu}$ as shown on~\fref{fig4} in the case of the
first layer.

The values of $\mu_c$ and $\beta_1$
that we consider suitable are the ones which allow an oscillating behaviour for 
the widest
interval in the variable $\ln t^{-\nu}$. A modification of the boundary 
exponent
$\beta_1$ would change the average slope of the   oscillating regime. This
could be due to corrections to scaling, but, if such corrections really 
existed,
they  should cancel in this range of temperatures (in the oscillating regime, 
$t$
goes to values as small as  $10^{-9}$). The other parameter, $\mu_c$, modifies
the number of oscillations  and we have choosen a
value leading to the largest number of such oscillations. A poor determination 
of the critical point $\mu_c'=\mu_c+\Delta\mu_c$ would indeed artificially 
introduce a correction to scaling term, since 
$t^\beta=(t'+\Delta\mu_c)^\beta\sim 
t'^\beta\left(1+\beta\frac{\Delta\mu_c}{t'}\right)$.

\begin{table}
\caption{Numerical values of the critical temperature and the magnetic
exponents for the surface and  bulk magnetizations. The figure in
brackets gives the uncertainty on the last digit.}
\footnotesize\rm
\begin{tabular}{@{}llllllll}
\br
&&\centre{3}{surface}&&\centre{2}{bulk}\\
\ns
&&\crule{3}&\quad&\crule{2}\\
$r$ &  $\theta_c$ 						& $\beta_1$ 		
	& $\beta_L$&$\beta_s$&& $\beta_{(n-1)}$  & $\beta_b$ \\
\mr
.1  & .963977634341 (5) &  1.00036 (2) &		\full			
&	.0002	(2)	&& .500087 (1) & .5002 (2)\\
.2  & .93187679929 (2)  &  1.00146 (2) &1.0015 (1)&		\full	   	
&& .50033 (1)  &\full \\
 .3 & .90314503363 (2)  &  1.0034 (1)  &1.0034 (1)&		\full		 
  && .50072 (6)  &\full \\
 .5 & .85404149087 (2)  &  1.0092 (1)	 &		\full			
&	.0094 (2)	&& .50187 (1)  & .505 (1)
\\
 .8 & .796437160887 (5)	&  1.02214 (2) &		\full			
&	.0216 (2)	&& .50419 (2)	 & \full\\
 1. & .76600595095 (2)  &  1.0327 (1)  &		\full			
&	.0302 (2)	&& .505777 (2) & .516
(1)\\
 1.5& .70902241601 (2)  &  1.0621 (1)	 &		\full			
&		\full		   && .50943 (1)  & \full\\
 2. & .67010909237 (2)  &  1.0913 (1)	 &		\full			
&	.087 (1)		&& .51186 (3)  & .538
(1)\\
 2.5& .64234629279 (2)  &  1.1178 (1)		&		\full			
&		\full	   	&&	.51294 (1)		& \full\\
 3. & .621796760462 (5) &  1.1410 (1)  &1.1410 (1)&	.133 (1)		
&& .5132 (1)   & .555 (1)
\\   
 3.5&	.60610567508 (2)	 &  1.1602 (1)		&		\full		
	&		\full	    &&	.51327 (4)		& \full\\
 4. & .593804120472 (5) &  1.1766 (1)  &		\full			
&	.1692 (4)	&& .5130 (1)   & .563
(1)\\   
 4.5&	.58394117369 (2)  &  1.1904 (1)		&		\full			
&		\full	   	&&	.5122 (1)			& 
\full\\
 5. & .5758805295248 (5)&  1.2026 (1)  &		\full			
&	.195 (1)		&& .51125 (2)  & .567
(1)\\   \br
\end{tabular}
\label{table1}
\end{table}

The corresponding values of $\theta_c$, $\beta_1$ and $\beta_{(n-1)}$ are given 
for
several values of the perturbation amplitude $r$ in \tref{table1}. The critical
exponent associated to the right surface ($m_L$) of the Fibonacci chain has 
also
been computed for different values of $r$ for the largest chain size. It gives, 
with a good accuracy,
the same value as for the left surface ($m_1$) as it can be seen by inspection 
in the table. The aperiodic sequence is indeed the same, seen from both ends, 
if 
we forget the last two digits.

Furthermore, the profiles of
\fref{fig1} clearly show that  the sites of the chain are not all equivalent 
and 
the
magnetization profiles can be locally rescaled with different values of the 
exponents depending
on the site~\cite{berche96}. Thus, after the local quantities, the computation 
of the surface
and mean bulk magnetizations enable us to determine the critical exponents 
respectively  written
$\beta_s$ and $\beta_b$ and given in \tref{table1}.

From our values, one obviously recovers the usual unperturbed ordinary 
transition values of
the exponents when the perturbation amplitude goes to zero.

\subsection{Susceptibility and critical isotherm} 

Taking account of a non vanishing bulk  magnetic field in equations~\eref{eq-8} 
and (7), 
one can compute the magnetization in a finite field  and then deduce 
  the critical isotherms exponents $\delta_{(n-1)}$ and $\delta_1$ 
from the behaviours of the local magnetizations $m_{(n-1)}$ and $m_1$ with 
respect to $h$:
\begin{equation}
m_{(n-1)}\sim h^{1/\delta_{(n-1)}},\qquad m_1\sim h^{1/\delta_1},\qquad t=0.
\label{eq-26}\end{equation}

\begin{figure}
\epsfxsize=14cm
\begin{center}
\mbox{\epsfbox{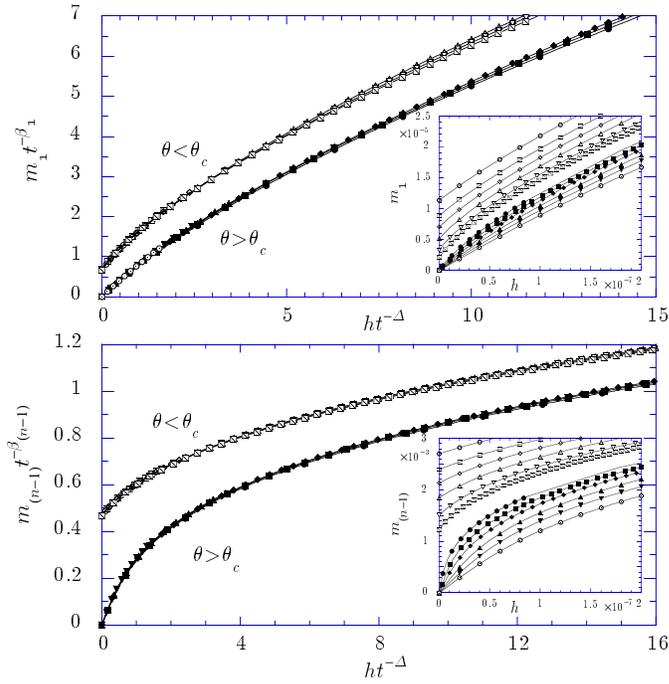}}
\end{center}
\vskip -0mm
\caption{Rescaled equations of state for the boundary and bulk magnetization 
for 
$r=2$. The
values of the temperature are $\theta=0.670090$, $0.670094$, $0.670097$,
$0.670100$, $0.670103$, $0.670105$ below $\theta_c$ and $0.670111$, $0.670113$,
$0.670115$, $0.670118$, $0.670121$, $0.670125$ above $\theta_c$. Top: scaling
functions $f_{m_1}^\pm$, the insert shows the boundary magnetization as a 
function of the bulk
magnetic field.  Bottom: same as above for the local bulk magnetization.} 
\label{fig5}   \end{figure}

This time, a direct log-log plot allows a precise determination of the
exponents and the rescaled equation of state confirms the
validity of the estimate since we obtain a good data collapse. In the case of 
the boundary
magnetization, the scaling assumption takes the following form under rescaling 
by an arbitrary factor
$b$:  \begin{equation} m_1(t,h)=b^{-\beta_1/\nu} m_1(b^{y_t}t,b^{y_h}h),
\label{eq-27}\end{equation}
where $y_t$ is given by the inverse of the correlation length exponent 
$y_t=1/\nu$ and the
value of the magnetic field anomalous dimension $y_h$ follows the requirements
of~\eref{eq-26}: $y_h=\beta_1\delta_1/\nu=\beta\delta/\nu$. The choice
$b=t^{-\nu}$ for the rescaling factor then
leads to a universal behaviour expressed in terms of a single scaled variable:
\begin{equation} m_1(t,h)=t^{\beta_1}f_{m_1}^\pm(ht^{-\Delta})
\label{eq-28}\end{equation}
where $\Delta=\beta_1\delta_1$ is the so-called gap exponent, $f_{m_1}^\pm$ is 
a
universal scaling function and $\pm$ refers to the two phases $\theta 
>\theta_c$ 
and $\theta <\theta_c$.
This may then be checked by a plot of $m_1t^{-\beta_1}$ v.s. $ht^{-\Delta}$
shown on~\fref{fig5} and the same type of universal function have been obtained
for the local bulk site $m_{(n-1)}t^{-\beta_{(n-1)}}=f_{m_{(n-1)}}^\pm 
(ht^{-\Delta})$. The values of
$\delta_1$ and $\delta_{(n-1)}$ are given in \tref{table2}.

\begin{table}
\caption{Numerical values of the critical exponents associated to the critical 
isotherms and
the susceptibilities. $\gamma_b$ and $ \delta_b$ correspond to the behaviour of 
the
mean bulk magnetization $ m_b$. The figure in brackets gives the uncertainty on 
the last
digit.}\footnotesize\rm
\begin{tabular}{@{}llllllllll}
 \br &\centre{4}{surface}&&\centre{4}{bulk}\\
\ns
&\crule{4}&\quad&\crule{4}\\
$r$& $\gamma_1$
&$\delta_1$&$\gamma_s$&$\delta_s$&&$\gamma_{(n-1)}$&$\delta_{(n-1)}$&$\gamma_b$
&
$\delta_b$ 
\\ \mr .1 &  .5013 (2) &1.5024 (2)&1.498 (1)	&		\full		
&&.9997 (1)&	2.9989 (1) &1.0005 (1)&\full\\
.2 &  .5006 (2) &1.5004 (2)&		\full			&		
\full		&&.9993 (2)&2.9972 (3)  &  \full  &\full\\
.3 &  .4992 (2) &1.4977 (2)&		\full			&			
\full			&&.9993 (2)&2.9949 (9)  & \full    &\full\\
.5 &  .4958 (2) &1.4901 (2)&1.493 (1)	&	312 (11)	&&.9989 
(2)&2.9895 (9)  &.99790 (2)&2.98136
(2) \\ 
.8 &  .487 (1)  &1.4751 (3)&1.486 (1)	&	85 (2)			&&.9986 
(3)&2.981 (2)   & \full   &\full\\
1. &  .4796 (2) &1.4641 (2)&1.480 (2)	&	53 (1)			&&.9985 
(4)&2.9744 (9)  &.99253 (2)&2.93144
(2)\\ 
1.5& .4568 (2)  &1.4378 (4)&  \full   &	\full			&&.9988 
(7)&2.963 (3)   &  \full   &\full\\
2. & .4316 (2)  &1.4135 (1)&1.438 (2) &	16.37 (2)&&.999  (2)&2.9571 (1)  &.9792 
(1) &2.82375
(3) \\ 
2.5& .412 (1)   &1.3845 (6)&		\full			&	\full		
&&.9992 (9)&2.954 (2)	  & \full   &\full\\ 
3. & .388 (1)   &1.3484 (6)&1.394 (2)	&	11.2 (2)	&&.9988 
(6)&2.952 (2)   &.9660 (1)
&2.7513 (2)\\   
3.5&	.372 (1)   &1.3108 (5)&  \full  	&	\full			
&&.9986 (5)&2.949 (2)	 & \full    &	\full			\\ 
4. & .354 (1)   &1.2989 (4)&1.360 (2)	&	10.01 (5)&&.9988 (8)&2.948 (2)  
&.9619 (5) &2.6976 (3)\\    
4.5&	.341 (1)   &1.2571 (2)&		\full			&	\full		
		&&.999  (1)&2.950 (2)	  & \full    &	\full\\ 
5. & .328 (1)   &1.2467 (6)&1.330 (2)	&	8.3 (4)		&&.999  
(2)&2.953 (2)   &.9514 (2) &2.65938
(2)\\   \br\end{tabular}
\label{table2}
\end{table}

The behaviours of $m_s$ and $ m_b$ with $h$ at the critical point lead to the
values of $\delta_s$ and $\delta_b$, also listed in \tref{table2}. We can point 
out the
low accuracy in the determination of $\delta_s$ since the slope of the log-log 
plot of $m_s$
v.s. $h$ is quite small when $r$ reaches the unperturbed value $r=0$.

\begin{figure}
\epsfxsize=11cm
\begin{center}
\mbox{\epsfbox{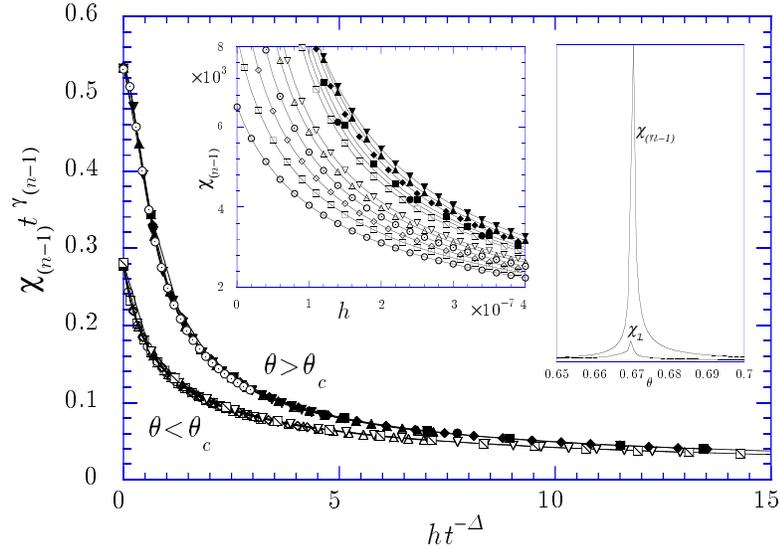}}
\end{center}
\vskip -0mm
\caption{Rescaled  bulk susceptibility giving the behaviour of the 
universal functions
$f_{\chi_{(n-1)}}^\pm$ below and above $\theta_c$  for $r=2$. The values of the
temperature are the same as in figure 6. The inserts show the
behaviours of $\chi_{(n-1)}$ as a function of $h$ for the same temperatures 
(left),
and the singularities of both $\chi_{(n-1)}$ and $\chi_1$ in zero magnetic 
field 
as a
function of $\theta$ (right).}  \label{fig6}   \end{figure}

The derivative of equation~\eref{eq-27} with respect to the bulk magnetic
field $h$  defines the boundary susceptibility $\chi_1$
which diverges as the critical point is approached with an exponent $\gamma_1$.
Numerically, the boundary  magnetization is calculated for several values of 
the 
bulk
magnetic field (of the order of $10^{-9}$), and $\chi_1$ follows a finite 
difference
derivation. The bulk local susceptibility $\chi_{(n-1)}$  may be obtained in
the same way. Log-periodic oscillations also occur in these quantities and the 
determination
of the exponents can be done in the same  way as in the previous section for 
the 
magnetization.
Again, the accuracy of the result is confirmed by the rescaled curves for the
susceptibilities, for example 
$\chi_{(n-1})t^{\gamma_{(n-1)}}=f_{\chi_{(n-1)}}^\pm (ht^{-\Delta})$
shown on \fref{fig6} exhibits a good data collapse on two universal curves for 
$\theta <\theta_c$ and
$\theta>\theta_c$.

 The values of the exponents are given in~\tref{table2} which presents also 
$\gamma_s$ and
$\gamma_b$, associated to the surface and average bulk magnetization field
derivatives.

\section{Specific heat}
\label{sec:therm}

According to the definitions of \sref{sec:GL},
the surface and bulk free energies are also defined as follows:
\numparts
\begin{eqnarray}
F_s={1\over 2}(F_{FBC}-F_{PBC}),\\
F_b=F_{PBC}.
\end{eqnarray}\label{eq-38}
\endnumparts
where $F_{FBC}$ and $F_{PBC}$  denote the total free energies of aperiodic 
chains
with free and periodic boundary conditions respectively and are obtained 
numerically using
equations \eref{eq-40} and \eref{eq-37}.

 The expected  singular behaviours of the free energy densities
\numparts
\begin{eqnarray}
f_s&(t,h)=t^{2-\alpha_s}f_s(ht^{-\Delta}),\\
f_b&(t,h)=t^{2-\alpha_b}f_b(ht^{-\Delta}),
\end{eqnarray}\label{eq-33}
\endnumparts
where the dependence of $f_s$ with the local magnetic surface field $h_1$ has 
been omitted
since we always consider the case $h_1=0$, lead to the surface and bulk 
specific 
heat
exponents.
 The values of $\alpha_s$ and $\alpha_b$  are simply deduced from the slopes of 
the log-log
plots of $f_s$ and $f_b$ v.s. $t$.

\begin{figure}
\epsfxsize=11cm
\begin{center}
\mbox{\epsfbox{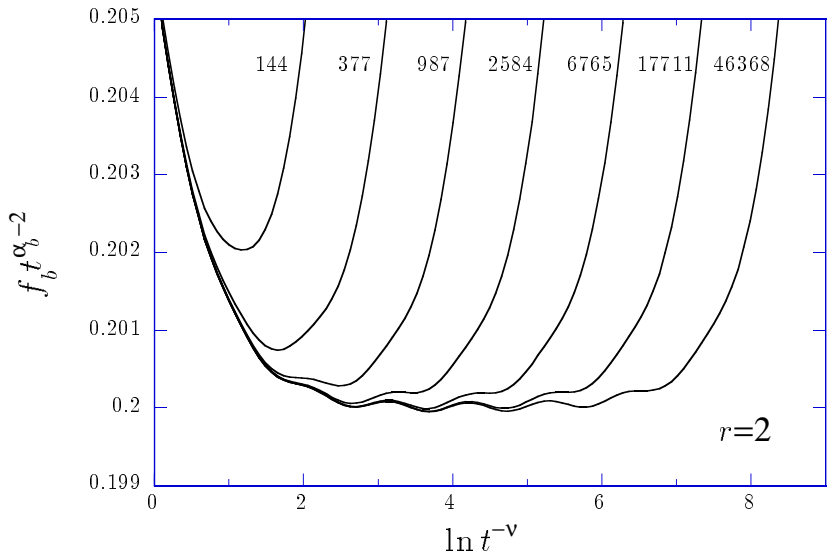}}
\end{center}
\vskip -0mm
\caption{Rescaled bulk free energy density $f_bt^{\alpha_b-2}$ v.s. $\ln 
t^{-\nu}$ for $r=2$. The amplitude of the bulk free energy density exhibits 
log-periodic oscillations.}  \label{fig20}   
\end{figure}

In \fref{fig20}, we show the bulk free energy density amplitude 
$f_bt^{\alpha_b-2}$ as a function of $\ln t^{-\nu}$ for $r=2$. It exhibits the 
same type of oscillating behaviour than the rescaled magnetisation of 
\fref{fig4}.

The surface and bulk specific heat exponents  are collected in \tref{table3}. 
The bulk 
specific heat discontinuity of the homogeneous system is washed out in the 
perturbed system, since $\alpha_b<0$.

\begin{table}
\caption{Numerical values of the specific heat critical exponents. The figure 
in
brackets gives the uncertainty on the last digit.}
\begin{indented}
\item[]\begin{tabular}{@{}lll}
\br
$r$ &  $\alpha_s$ & $\alpha_b$   \\
\mr
.1&0.51496 (7) &-0.00031 (1) \\
.5&0.50112 (5) &-0.00733 (1) \\
.8&0.48448 (5) &-0.01709 (1) \\
1.&0.47075 (4) &-0.02462 (1) \\
2.&0.40265 (1) &-0.05924 (1) \\
3.&0.35077 (4) &-0.07813 (1) \\
4.&0.31516 (3) &-0.08579 (1) \\
5.&0.28935 (6) &-0.08805 (1) \\
\br
\end{tabular}
\end{indented}
\label{table3}
\end{table}

\section{Discussion}
\label{sec:concl}

\begin{figure}
\epsfxsize=14cm
\begin{center}
\mbox{\epsfbox{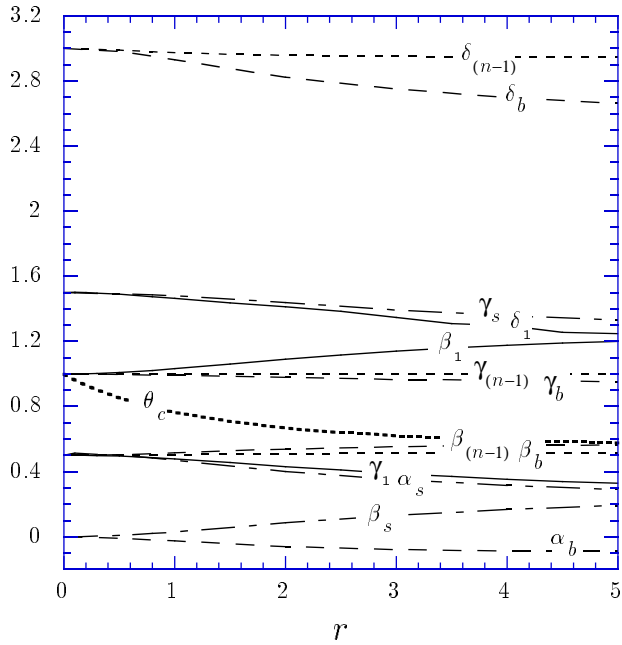}}
\end{center}
\vskip -0mm
\caption{Variations of the surface and bulk exponents with the
 perturbation amplitude $r$ (boundary exponents: \full, surface: \chain, local
bulk: \dashed, mean bulk: \broken).}  \label{fig9}   
\end{figure}

We have calculated numerically several surface and bulk critical exponents for 
a 
marginal
aperiodic system within  mean field theory. The marginal aperiodicity leads to
exponents which vary continuously  with the amplitude of the perturbation $r$. 
The variations
of these exponents are shown on \fref{fig9} as a function of $r$. 

The comparison in \tref{table1} between the bulk exponent 
$\beta_b$ and the local
one $\beta_{(n-1)}$ clearly shows that it is no longer possible, in this 
aperiodic system, to define a
unique bulk exponent, as it was already suggested by the possibility of a local 
rescaling of the
profiles with  position-dependent exponents which suggests a multiscaling 
behaviour. 
A constant value
$y_t=1/\nu$ is consistent with continuously varying exponents,  in order 
to keep a
vanishing crossover exponent which ensures that the marginality condition 
remains valid for any value
of the aperiodicity amplitude $r$. For $y_h$ on the other hand, there is no 
such 
reason. 
From this point of 
view, equations like
\eref{eq-27} are not exact since a unique field anomalous dimension $y_h$ has 
no 
real significance. It follows that the universal  functions  in 
\fref{fig5} and \fref{fig6} only give an approximate picture of the scaling 
behaviour in this system, since they involve the gap exponent
$\Delta=y_h/y_t$. The good data collapse has to be credited to the weak 
variation of the exponents with the perturbation amplitude $r$.

On the other hand, the scaling laws involving the dimension of the system are 
satisfied in mean field theory with  a value of $d$ equal to the upper critical 
dimension $d^*$. As for the $2d$ Ising model with a marginal 
aperiodicity~\cite{berche95,berche96}, one expects a strongly anisotropic 
behaviour in the Gaussian model. 
It yields a continuous shift of the upper critical dimension with the 
pertubation amplitude, $d^*(r)$, 
since the value $d^*=4$ for a critical point in the homogeneous $\phi^4$ theory 
follows  Ginzburg's criterion for an isotropic  behaviour. Hyperscaling 
relations should thus be satisfied for the mean field exponents 
 with $d^*(r)$:
\numparts
\begin{eqnarray}
2-\alpha_b=\nu d^*(r),\\
2-\alpha_s=\nu (d^*(r)-1).
\end{eqnarray}\label{eq-50}
\endnumparts
We can make use of these relations to obtain an estimate of the upper critical 
dimension $d^*(r)$ for this aperiodic system. The corresponding results are 
given in \tref{table10}.

\begin{table}
\caption{Numerical values of the upper critical dimension $d^*(r)$ deduced 
from hyperscaling relations.}
\begin{indented}
\item[]\begin{tabular}{@{}ccc}
\br
$r$ &  $(2-\alpha_b)/\nu$ & $(2-\alpha_s)/\nu+1$   \\
\mr
.1&4.00 &3.97 \\
.5&4.01 &4.00 \\
.8&4.03 &4.03 \\
1.&4.05 &4.06 \\
2.&4.12 &4.19 \\
3.&4.16 &4.28 \\
4.&4.17 &4.37 \\
5.&4.18 &4.42 \\
\br
\end{tabular}
\end{indented}
\label{table10}
\end{table}

The two determinations are in good agreement for small values of the 
perturbation amplitude. The discrepancy at larger values of $r$
 suggests that the precision in the determination of the exponents has 
probably been
overestimated, but 
the variation of the upper critical dimension with the perturbation amplitude 
is 
clear and should be attributed 
to an anisotropic scaling behaviour in the corresponding Gaussian model. 

One can finally mention that a mean field approach for relevant aperiodic 
perturbations would be interesting. Many cases of aperiodic sequences with a
wandering exponent $\omega>-1$ are known, they constitute  relevant 
perturbations in mean field theory. In the case of the $2d$ layered Ising model 
with relevant perturbations,
a behaviour which looks like random systems behaviour, with essential 
singularities, was found~\cite{iglotu94}, and the same type of situation can be 
expected within mean field approximation.

\ack 
We thank L. Turban  for valuable discussions and F. Igl\'oi and G. Pal\`agyi 
for 
informing us on a
related work before publication. This work has been supported by the Groupe 
CNI/Mati\`ere under
project CNI155C96.

\end{document}